\documentclass{ieee}
\usepackage{times}

\usepackage{graphicx}
\newtheorem{def.}{Definition}
\newtheorem{the.}{Theorem}
\newtheorem{pro.}{Proof}

\begin{document}
\title{A Practical Approach for Circuit Routing on Dynamic Reconfigurable Devices\thanks{Supported in part by the German Science Foundation (DFG), SPP~1148 (Rekonfigurierbare Rechensysteme).}}
\author{
Ali Ahmadinia, Christophe Bobda, Ji Ding, Mateusz Majer, J\"urgen Teich
\\Department of Computer Science 12 
\\University of Erlangen-Nuremberg, Germany\\ 
\{ahmadinia, bobda, mateusz, teich\}@cs.fau.de
\and
S\'andor P.\ Fekete, Jan C.\ van der Veen
\\Department of Mathematical Optimization 
\\Braunschweig University of Technology, Germany
\\ \{s.fekete, j.van-der-veen\}@tu-bs.de\\
}
\maketitle
\thispagestyle{empty}
\begin{abstract}
Management of communication by on-line routing in new FPGAs with a large amount of logic resources and partial reconfigurability is a new challenging problem. A Network-on-Chip (NoC) typically uses packet routing mechanism, which has often unsafe data transfers, and network interface overhead. In this paper, circuit routing for such dynamic NoCs is investigated, and a practical 1-dimensional network with an efficient routing algorithm is proposed and implemented. Also, this concept has been extended to the 2-dimensional case. The implementation results show the low area overhead and high performance of this network.
\end{abstract}
\section{Introduction} \label{intro}

The amount of logic resources in FPGAs is growing continuously and their dynamic configuration abilities lead us to multitasking systems, which need resource and communication management. Different on-line placement algorithms as central part of resource management have been proposed and developed \cite{ABFTV04}\cite{baz}. However, the communication management is sitll challenging. In \cite{ABFTV04}, the routing cost is considered during placement; however, this approach does not give any routing algorithm and structure to solve it. The issue of communication management has been referred to a Network-on-Chip (NoC), which is an emerging research topic nowadays.\\
Networks-on-Chip have been shown to be a good solution to support communication on System-on-Chip. Using an on-chip interconnection network to replace top-level global routing has the advantages of structure, performance, and modularity. A chip employing a NoC is composed of a set of network clients like DSP, memory, peripheral controller, custom logic, etc. Most of the existing work proposed today uses packet routing for communication between modules \cite{vernalde} \cite{dynoc}. However, packet routing has two main disadvantages: First, each module has an area overhead, because it needs a network interface to split data into packets at the source and merge them at the destination. Second, it reduces the performance, since if one of the packets is lost, the destination cannot use the transferred data and the lost packet must be sent again. The other possibility is circuit routing, which establishes a physical connection between source and destination by setting required switches. Therefore, compared to circuit routing, packet-based approaches use routing resources more efficiently by sharing them for different connections, but on the other hand, they have a network interface overhead, and low performance data transfer.\\
In this paper, circuit routing for NoCs is investigated, and for a modified topology from multi-processor network (Reconfigurable Multiple Bus) \cite{elgindy}, a local and efficient circuit routing algorithm is presented. This topology has been already developed as a ring topology with packet switching in multicomputer systems, which we adapted to 1-dimensional NoCs with circuit switching. We also extended the network structure and the routing algorithm to the 2-dimensional case.\\
The rest of the paper is organized as follows: in Section \ref{arch}, we review briefly existing on-chip network infrastructures. Section \ref{RMB} deals with the Reconfigurable Multiple Bus on Chip (RMBoC) structure, and our routing approach. The extension of RMBoC to 2-dimensional networks is presented in Section \ref{XRMB}. Section \ref{impl} contains the details of our implementation, and the restrictions and the challenges of implementation for using the infrastructure with circuit routing in the case of partial reconfiguration is presented in Section \ref{DRC}. The results are given in Section \ref{res}, and Section \ref{concl} concludes the work and suggests future work.
\section{Interconnection Network Architectures} \label{arch}
The choice of the system-level communication architecture has a significant impact on system performance and energy consumption. We give a short overview of some proposed interconnection structures for on-chip networks.\\
The best-known infrastructure is the bus architecture. The Advanced Microcontroller Bus Architecture (AMBA) from ARM, the CoreConnect from IBM, and WISHBONE from Silicore are some existing bus-based communication architectures for SoCs. Traditionally, they have been used for data path interconnection because of their simplicity. However, only one module can drive the network at a time. Moreover, a bus arbiter is needed when several processors attempt to use the bus simultaneously. As a result, all connections must be determined by the arbiter, and then the routing approach has low performance, and is not scalable \cite{micheli}.\\
A mesh-based interconnection network has been suggested for System-on-Chip in \cite{vernalde}, where an array of routers interconnects an array of processors. The router network has a 2-dimensional torus topology to limit hardware overhead. It has been implemented in a 1-dimensional structure, a wormhole routing (which is a packet routing) is adopted.\\
Dehon et. al. \cite{dehon} have proposed a Fat-Tree topology for an on-chip interconnection network.
There is a unique set of switches between any source and sink in this network. For finding the routes, it needs still global approaches, and the pathfinder algorithm has been used \cite{pathfinder}.\\
Another topology that has been used for NoCs is a hexagonal mesh or Honeycomb \cite{jantsch}. Each resource is directly connected to three switches and can reach 12 resources with a single hop. The main advantages of this topology are that fewer hops are needed for connecting resources, and the ratio of resources to the switches being three.\\ 
With the exception of the Fat Tree structure, all of the above architectures have been applied to packet switching. The Fat Tree topology needs a global routing algorithm for establishing the connections by finding the shortest path, and then reducing congestion of shared segments.\\
For reason of highest speed, we want to develop a circuit routing for interconnections on chip, which has the following features and advantages:
\begin{itemize}
\item The infrastructure, including switches and their connections, occupies a small area (low area overhead).
\item The routing connections can be determined fast and locally at switches.
\end{itemize}
For achieving the mentioned features, we have chosen to use the concept of the Reconfigurable Multiple Bus(RMB) Network \cite{elgindy}, which is proposed for multi-processor networks. We have modified the RMB to use as a Network-on-Chip. This interconnection structure is called RMBoC, and explained in the next section. 
\section{RMBoC Structure} \label{RMB}
\begin{figure}[!htb]
  \centering
  \includegraphics[width=0.7\linewidth]{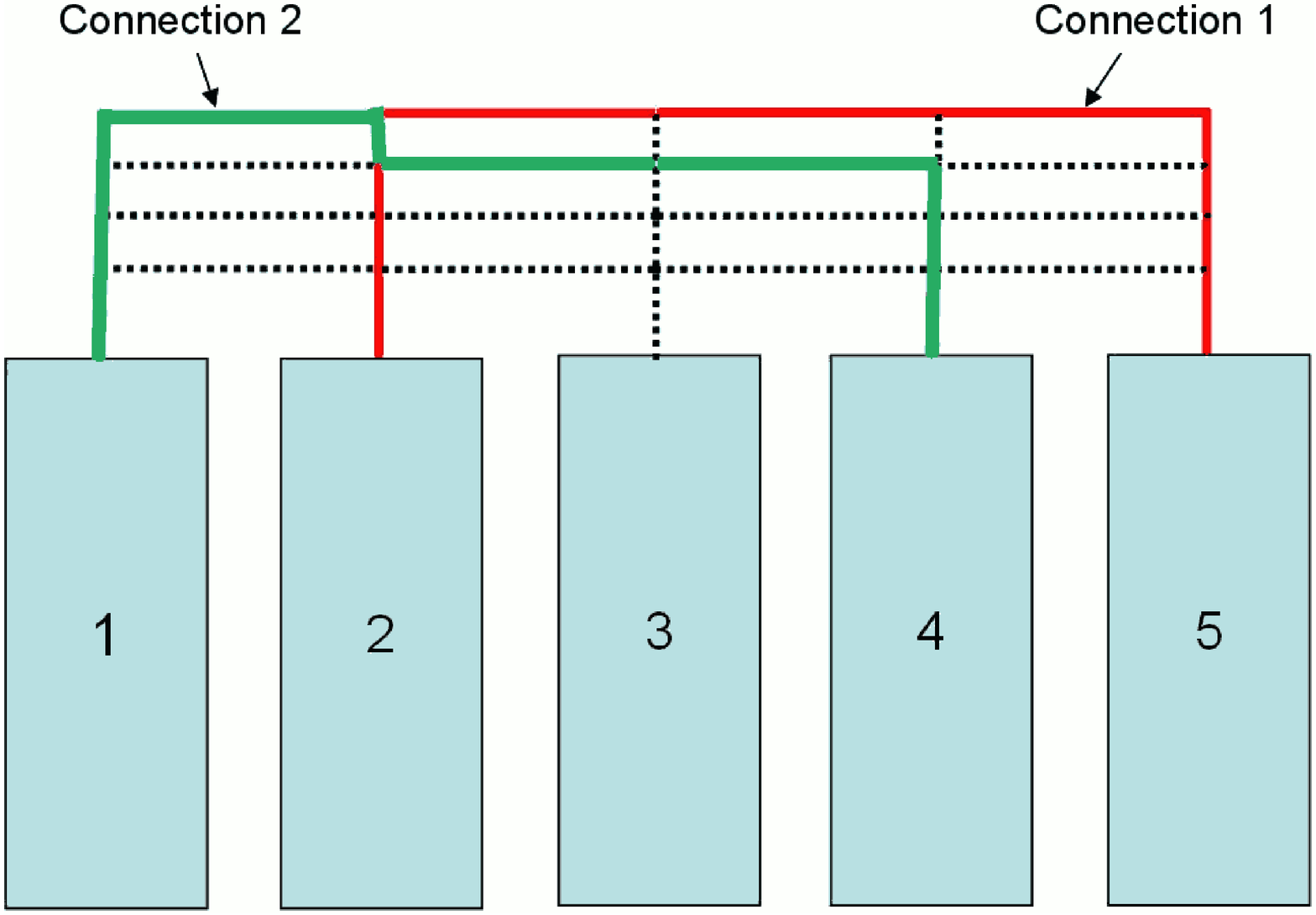}
  \caption[]{Routing Strategy on RMB.}
  \label{rmb-routing}
\end{figure}

\begin{figure}[!htb]
  \centering
  \includegraphics[width=0.7\linewidth]{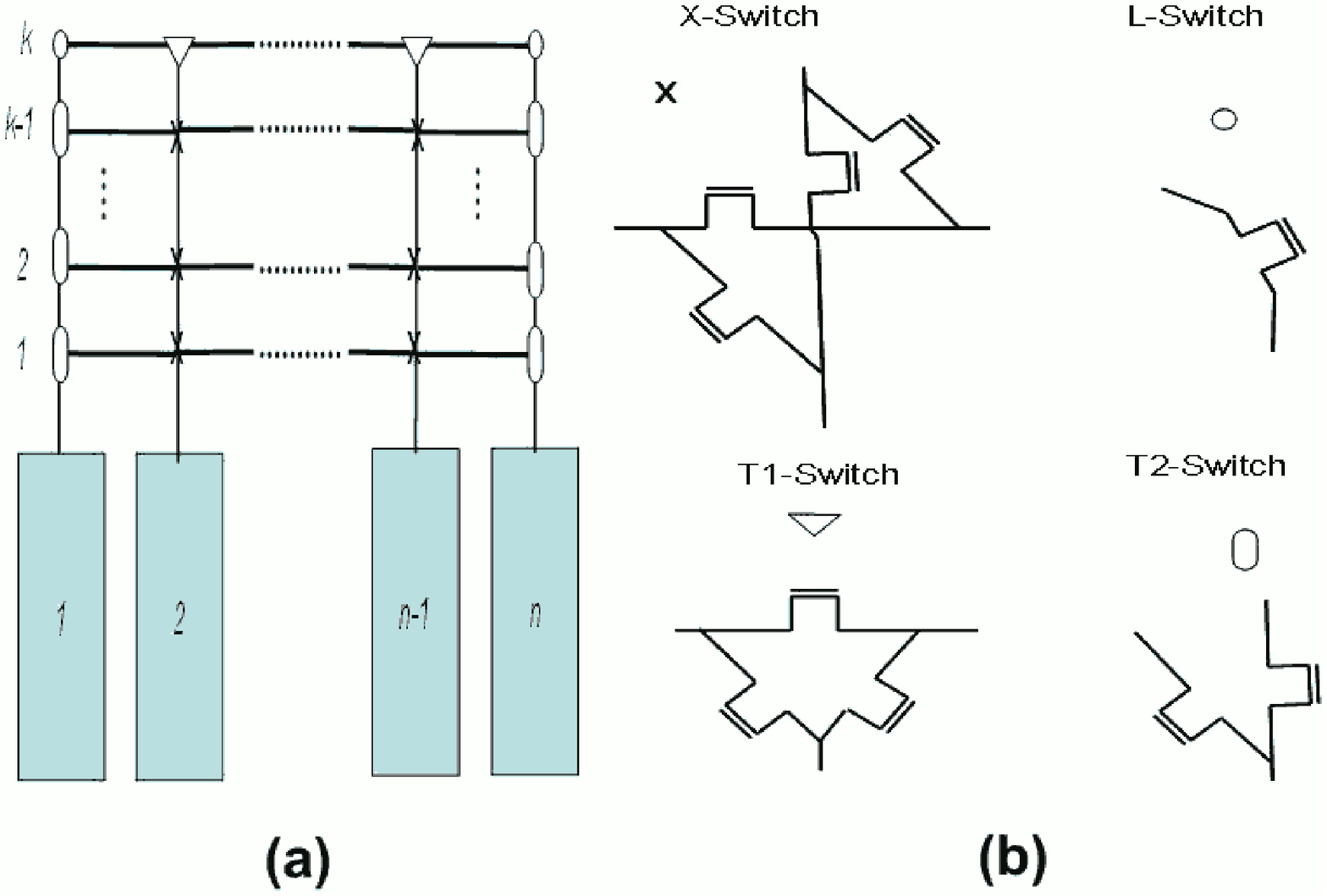}
	\caption{(a) RMBoC Architecture (b) Basic structure of switches in RMBoC.}
	\label{1DRMB}
\end{figure}

The reconfigurable multiple bus architecture relies on the use of an array of parallel bus segments between processing nodes. Each processing node can access the reconfigurable bus system to communicate with another processing node. The bus controller connected to each node coordinates the efficient use of available buses through reconfiguration. The most important aspect of this architecture is that the reconfiguration takes place entirely independently of any current communication in which the bus segments are involved \cite{elgindy}.\\
RMB as a ring-based topology has been proposed to implement a medium-size  multi-processor system. The processors send messages through RMB using a mechanism based on wormhole routing. New channels of communication are allocated at the top segments.\\
For example in Figure \ref{rmb-routing}, first by using highest free segments, a connection between modules 2 and 5 is established and then module 4 is routed to module 1 through highest free ones. During the lifetime of this communication, the allocated channel will be moved down to other free channels. This process is called bus compaction, which is used for reducing the establishment time of a connection.
\\We have changed three main aspects of this approach:
\begin{itemize}
\item Instead of having a ring-based topology, we use a 1-dimensional array. In order to implement a ring on an FPGA, global routing lines through the network would be required, which prevents dynamic reconfiguration or slow wrap-around connections outside of the FPGA would be needed.
\item We do not make any compaction for moving all occupied segments to the bottom free segments, because for compaction, signal assignment conflicts will happen.
\item We do not use any protocol for packet switching, and all the routings are circuit-switched and controlled by signaling.
\end{itemize}
An RMBoC with $n$ processing elements and $k$ buses is depicted in Figure \ref{1DRMB}. For establishing a connection between any two processors, the highest free bus segments will be selected dynamically. As shown in Figure \ref{1DRMB}, four types of switches have been used. The basic structure of these switches as depicted in Figure \ref{1DRMB} is very simple and uses very few transistors.\\

This network architecture is appropriate for Xilinx FPGAs that have a column-based configuration architecture and can be used as 1-dimensional networks for run-time reconfiguration. Details of this implementation are presented in Section \ref{impl}.
\section{Extension of RMBoC} \label{XRMB}
\begin{figure}[!htb]
  \centering
  \includegraphics[width=0.57\linewidth]{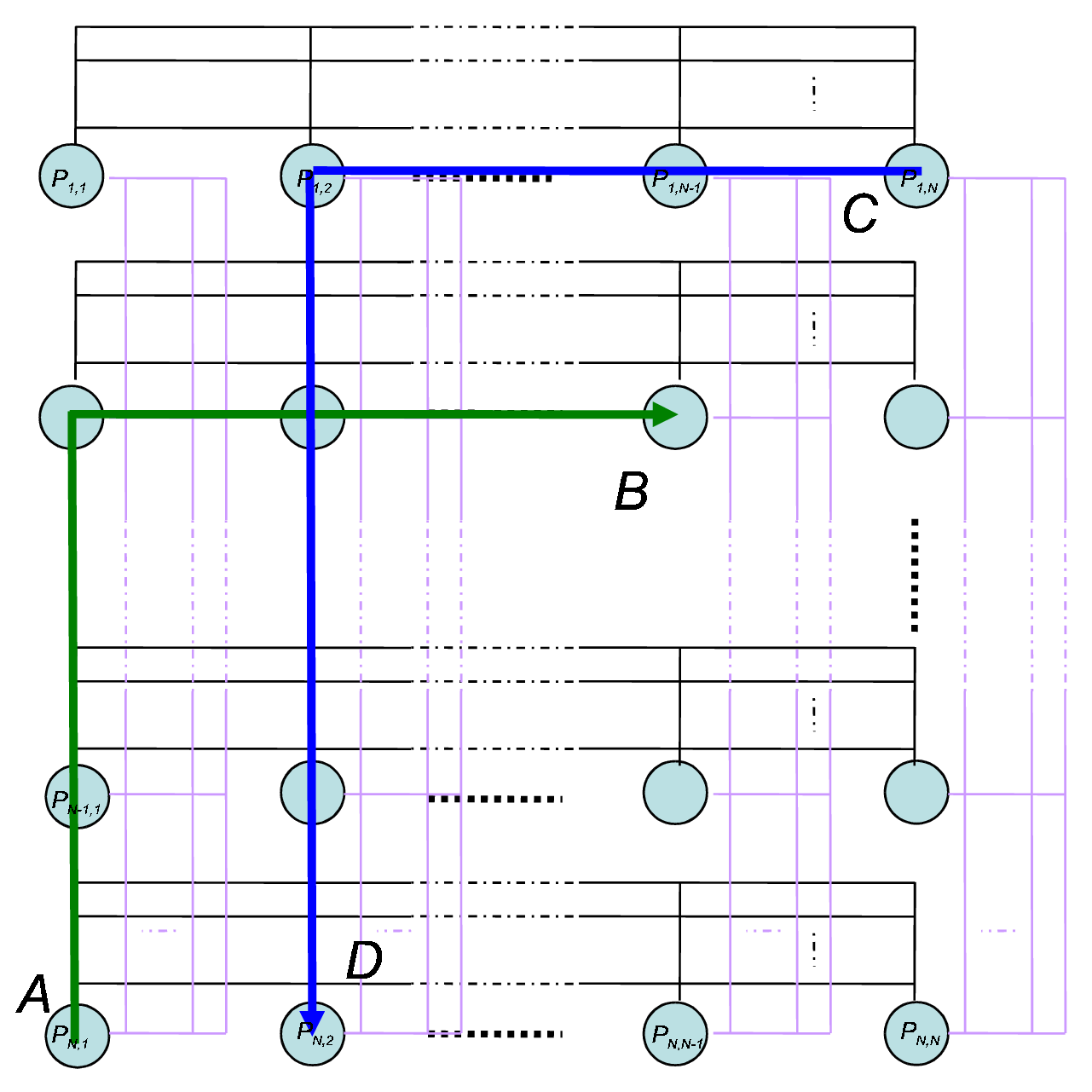}
  \caption[]{2-Dimensional RMBoC.}
  \label{2DRMB-routing}
\end{figure}

We have also extended the RMBoC to 2-dimensional networks. The main reasons for this extension are:
\begin{itemize}
\item To increase the utilization of FPGA resources, we need network architecture similar to popular FPGA architectures (mesh-based).
\item In order to realize a fully-connected 1-dimensional network with $n$ processors, $O(n^2)$ parallel buses are needed. For a fully-connected 2-dimensional network with $n=N\times N$ processors, $N\times O(N^2)=O(N^3)=O(n^{1.5})$ buses are required. Then, bus segments can be used more efficiently.
\end{itemize}
A 2-dimensional RMBoC with $N\times N$ processors and $k$ buses in each row and column is shown in Figure \ref{2DRMB-routing}. For establishing a route, the connections trend to go upward, i.e., upward is the first choice in each switch according to the destination location. If the destination is located at a lower level, the right and leftwards channels will be used, depending on the sink. Only when a route reaches the same column as destination and the destination is at a lower level, the downward channel is selected. For example, you can see in Figure \ref{2DRMB-routing} the routings from $A$ to $B$ and $C$ to $D$.
\section{Implementation} \label{impl}
In this section, we present the details of our implementation on Xilinx FPGAs that have a column-based configuration structure. We have focused on implementing the 1-dimensional RMBoC on these. Also, we have implemented the 2-dimensional model to analyze the characteristics and resource requirements of the network.\\
\begin{figure}[!htb]
  \centering
  \includegraphics[width=0.9\linewidth]{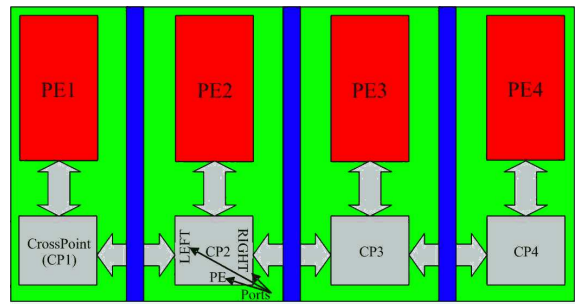}
  \caption[]{Implementation of the 1-D RMBoC.}
  \label{crs-pnt-chn}
\end{figure}
In this system, the actual crosspoints in one column are merged into one controller, which is different from the conceptual structure mentioned before (see Figure \ref{crs-pnt-chn}). If the separated crosspoint structure would be used, these points would have to communicate with each other to find out the free channel, which takes more clock cycles. However, if they are combined into one block, the decision can be done within one clock cycle. Furthermore, separate structures need more FIFOs for storing the unprocessed requests, while the combined one requires only one FIFO. Therefore, we call the combined switches in one column {\em crosspoints}.\\
In our example, the whole system consists of four modules; each one is a so-called crosspoint. Inside a single crosspoint, there are three kinds of structures: a) controller, b) data network and c) FIFOs, as shown in Figure \ref{crosspoint}. In the following, the function of these modules is explained:
\begin{figure}[!htb]
  \centering
  \includegraphics[width=\linewidth]{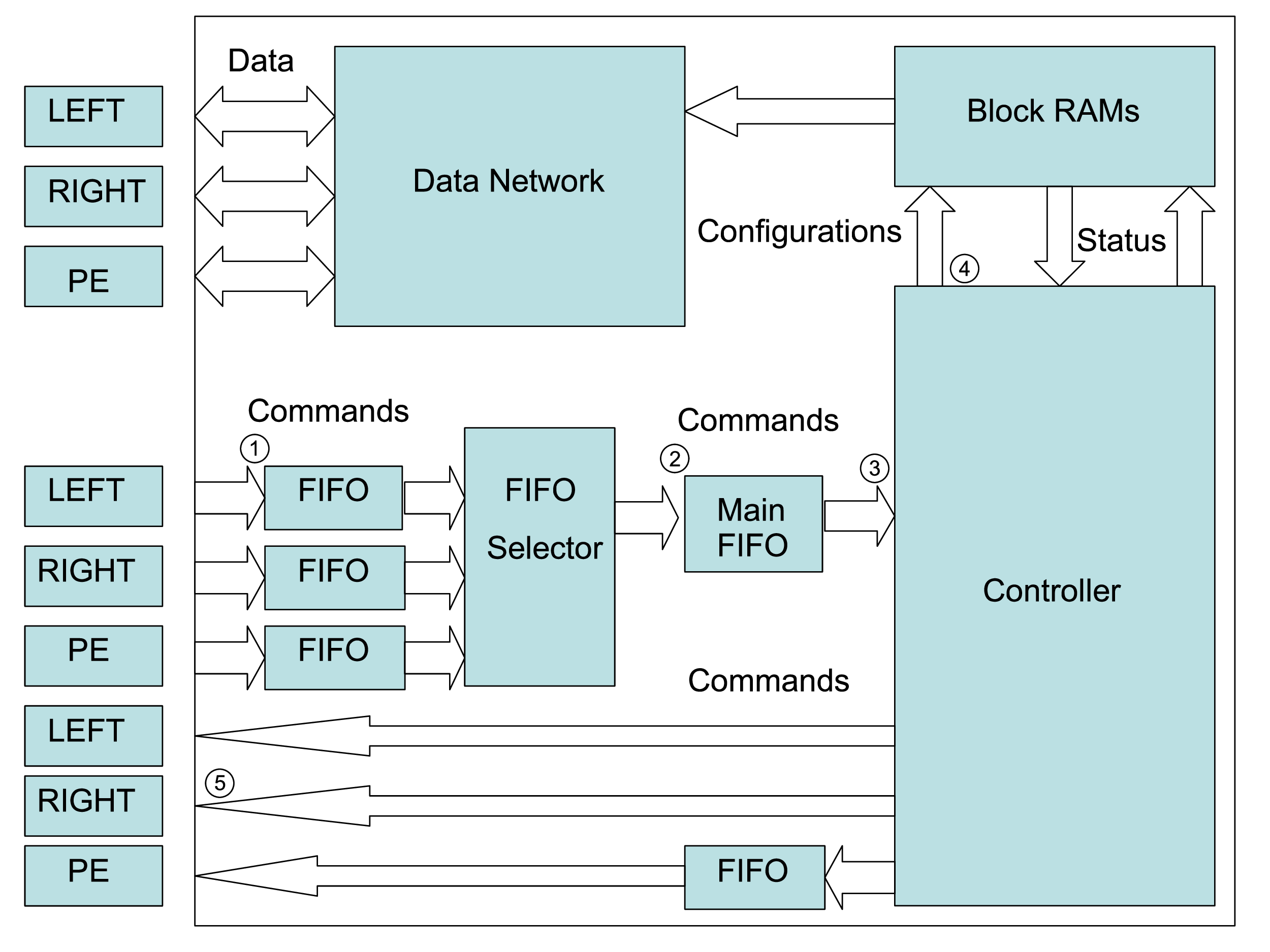}
  \caption[]{Architecture of a crosspoint.}
  \label{crosspoint}
\end{figure}
\\
\textbf{Controller:}  
The function of the controller is to transport control commands from one processor to another and to configure data channels between processors. In total, there are four kinds of commands: REQUEST, REPLY, CANCEL and DESTROY. The processors may use these commands in the following way: \\
First, one processor sends a REQUEST command to the corresponding crosspoint with the destination address. Then the crosspoint decides in which direction the command should be transferred and then written to the output buffer. During this period, no physical channel is created, because it is possible that this REQUEST is not be confirmed by the destination processor. This does not delay the connection establishment, because the data transfer from source cannot be started before getting the acknowledgment of all required segment allocations. When the next crosspoint gets the REQUEST, it will check the destination and then decide to transfer the command to its own corresponding processor or the next crosspoint.\\ 
When the destination is reached, the processor gets the request from the corresponding crosspoint and decides whether the channel can be created or not. If so, the REPLY command should be sent; if not, the CANCEL command should be sent. When a crosspoint gets a REPLY command, it will search for a free channel. If such a channel is available, then the configuration of the physical data network will be adjusted. If not, a DESTROY command will be sent back to the destination automatically to free all the previously created channels and also a CANCEL command will be sent to the source automatically to inform that the channel cannot be created. When the REPLY command reaches the source processor, then the complete channel will be created.\\ 
When a crosspoint gets a CANCEL command, it simply transfers it to the next crosspoint or processor. No modifications will be done to the configurations. After data transfer, the source processor wants to close the data channel, then the command DESTROY should be sent to the crosspoint. When the crosspoint gets the DESTROY command, it will loop up in its configuration registers, then destroy the corresponding channel and transfer the command to the processor or the next point.\\
By using the above protocol, different channels are allowed for each processor at the same time, as long as the channels are free. Channels with reversed source and destination are considered to be different channels.
The function of the mentioned commands of the controller are summarized as follows:\\
- REQUEST : To establish a connection.\\
- REPLY : Acceptance of the connection request by the destination.\\
- CANCEL : Rejection of the connection request by the destination.\\
- DESTROY: Deallocation of the occupied channels of the requested connection, when there is no free channel for establishing the complete path.\\
\textbf{Data network:}
The function of data network is just to connect corresponding data channels according to the configurations modified by the controller. Once the connection is established, data is transferred within one clock cycle from source to destination.
\\ \textbf{FIFOs:}
The purpose of the FIFOs is to provide buffer for commands. The FIFO selector sends the command from FIFOs of each side to the main FIFO. The policy of arbitration in the FIFO selector is Round-Robin (in order of Left, Right, and PE).\\
The reason that the main FIFO and a FIFO selector are used is that three function blocks would be needed for processing commands from left, 
right and bottom FIFOs. After processing the commands, some glue logic, which is needed to connect to the three blocks, has to be used to decide which block can be written to the output (3 to 1).\\
However, the three function blocks are similar, so they can be simplified to one block. The simplified block processes all the commands from left, right, and bottom. So a FIFO selector is needed to collect all the commands from different directions and store them in one extra FIFO.
Thus, the area of three function blocks is reduced to 1/3.\\ 
A single crosspoint consists of the above three structures, which should be connected to a processor. We have made measurements for a system consisting of four modules, i.e., four crosspoints and four processors. They are placed in parallel and connected by so-called bus-macros to enable partial reconfigurability.\\
In the top-level structure of a 2-dimensional RMBoC, there is a total of 16 processors, each of them connecting two crosspoints, one for row transfer and the other for column transfer. Consequently, 32 crosspoints are used, 16 for row connections and 16 for column connections. The main difference of crosspoints in one and two dimensions is the address width. As to the behavior of the processor, now they have to decide which crosspoint should be used, the one from row connection or from column connection. Another task of the processor is to transfer commands to other processors by switching from row connection to column connection, and vice versa.
\begin{figure}[!htb]
  \centering
  \includegraphics[width=0.9\linewidth]{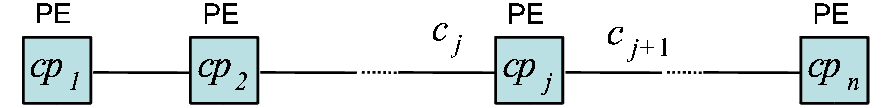}
  \caption[]{Processing of command $c_j$ in $cp_j$.}
  \label{req}
\end{figure}
To summarize the communication protocol, main steps of processing of a command $c_{j}\in \{REQUEST, REPLY, CANCEL, DESTROY\}$ in the crosspoint $cp_j$ to perform required operation(s) and generate command $c_{j+1}$ (Figure \ref{req}), is given as follows:\\
\textit{\\
1 Read the command $c_{j}$ from the left side FIFO\\
2 Write the command $c_{j}$ in the main FIFO\\
3 Read the command $c_{j}$ by the controller\\
4 Determine the new command $c_{j+1}$, and update the configuration of channels if needed\\
5 Write the command $c_{j+1}$ in the right side FIFO\\
}\\
Steps 1 to 3 each take two cycles, steps 4 and 5 execute in parallel, and need 2 cycles together. In the best case of processing a command, the FIFOs are empty, and in 8 cycles (for the 5 steps), the command will be processed.\\
To analyze the maximum delay for processing a command, some definitions are required:
\begin{def.} {\bf MaxTotalComm}
is the maximum number of commands in all three directions (Left, Right, Processing Element).
\end{def.}
\begin{the.}
The maximum required processing time is $(MaxTotalComm-1)\times 4 +4$ cycles.
\end{the.}
\textbf{Proof:}
\textit{The main steps will be executed in a pipeline model, in which steps 1 and 2 are in the first stage, and the rest in the second stage, therefore the waiting time would be four cycles for each command in the fifos, and four cycles for the second stage of the pipeline.}
\begin{the.}
The $MaxTotalComm$ is equal to $\lceil\frac{n^2+2n-4}{2}\rceil$.
\end{the.}
\textbf{Proof:}
\textit{To compute $MaxTotalComm$, the maximum number should be computed separately in each direction. For example in crosspoint $cp_j$ (Figure \ref{req}), the maximum number of requests from the right side would be that all the modules on the right of $cp_j$ send commands to all processing elements on the left of $cp_j$ including PE $j$ : $(n-j)\times j$. We can compute the maximum in similar ways for the other two directions. From the left side, at most $(j-1)\times (n-j+1)$ commands and from $PE$, at most $(n-1)$ commands can be received. As can be seen, the maximum numbers of requests from right and left sides are dependent on the position of crosspoint $cp_j$. It can be easily proved that these two numbers are maximized when $j=n/2$, and then $MaxTotalComm=\lceil\frac{n^2+2n-4}{2}\rceil$.\\}
It should be noted that this maximum delay for a command can happen only once through the network: When a command reaches its next crosspoint, and the commands ahead of it have already been processed, then no waiting cycle arises. Moreover, it has to be guaranteed that the number of commands in each direction is restricted to the depth of FIFOs; if the FIFO depth is smaller than the maximum number of simultaneous commands in a direction, then some commands can be lost, and the source has to send them again.
\section{Dynamic Reconfiguration Challenges} \label{DRC}
For enabling partial reconfiguration in the RMBoC structure, hard busmacros are used to fix the communications problem of crosspoints during reconfiguration. The busmacro ensures the reproducibility of the design routing and is implemented using
tri-state buffers. The tri-state buffers force the routing to always pass through the same places. At the same time they decouple the modules from each other during reconfiguration, avoiding possibly harmful transitory situations. In this
way, a 4 bit data bandwidth per row communication channel is possible between adjacent modules. This limitation comes from the current Virtex architecture and its limited routing resources.\\
Some problems should still be considered. For example in Figure \ref{crs-pnt-chn}, assume there are connections between $PE1$ and $PE3$, and also between $PE1$ and $PE4$. If now $PE2$ has to be reconfigured, what has to happen with the configuration of the $CP2$?\\
Virtex II (Pro) devices offer glitchless partial reconfiguration. If a configuration bit holds the same value before and after configuration, there will be no glitch on the resource that bit controls. Resources requiring special attention are SRL16s and LUT RAMs, because they change dynamically and will be overwritten when configuration occurs \cite{blodget}.\\
Therefore the data on the segments will be sent without any glitch. The only remaining problem concerns the state values of the crosspoint. Since these values will be lost during reconfiguration, if they are saved in LUTs, we use BlockRAMs that are distributed in six regions of the FPGA area. This means that we cannot use more than six modules in the RMBoC, and four modules for the appropriate structure. \\
The other problem that can arise with dynamic reconfiguration is loss of non-completed connection requests. For example in Figure \ref{crosspoint}, $PE4$ requests for a connection with $PE1$, and this request occupies a free segment in $CP3$ for this connection. Before allocating a free segment in $CP2$, and exactly at the same time when the request in $CP2$ is read from, $PE2$ will be reconfigured. Then the information of this request will be lost, and the whole request times out, because the source does not receive any acknowledgement from destination. The source sends the request again, but a free segment in $CP3$ is occupied uselessly because of the non-completed request. To solve this problem at each crosspoint, only one channel for requests with the same source and destination should be allocated.\\
Also by reconfiguration of modules, the DESTROY command may be lost, and the destruction of the connection will not be completed. Therefore, an additional command CONFIRM is added to acknowledge completion of connection destruction. Obviously if the source does not receive the CONFIRM from destination, it will initiate a new DESTROY command.


\section{Analysis Results} \label{res}
After the implementation, we compared the area overhead and performance of the RMBoC. As shown in Table \ref{res-1d}, the 1-dimensional RMBoC has been implemented on a Virtex II 6000 with four processors ($n=4$) and four parallel buses ($k=4$), 
 with a data bandwidth of 16 bits ($w=16$). The area overhead grows and the maximum frequency decreases with increasing data bandwidth. The area overhead range is relatively low (from 4\% to \%15 of FPGA area), and the reachable frequency is about 120 MHz.

\begin{figure}[!htb]
  \centering
  \includegraphics[width=\linewidth]{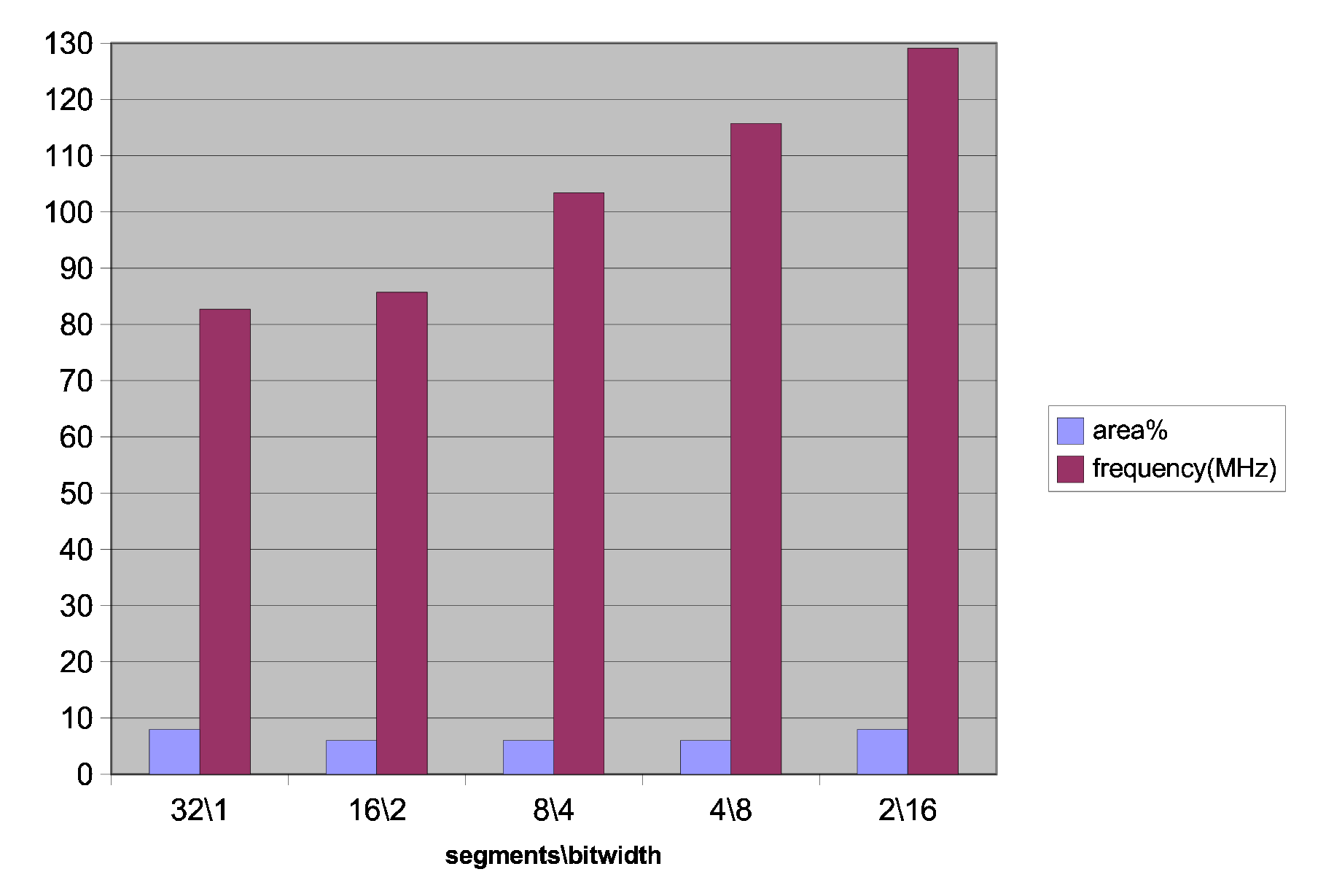}
  \caption[]{Tradeoff of number of segments $k$, and data bit width $w$ with $n=4$ fixed number of modules and $k\times w$ fixed maximal bandwidth.}
  \label{tradeoff}
\end{figure}

Also to analyze the behavior of our design by increasing the number of segments or data bitwidth, we have compared them such that the whole maximal bandwidth of the network ($k\times w$) is fixed (32). As depicted in Figure \ref{tradeoff}, by decreasing the number of segments $k$ and increasing the segment bitwidth $w$, the utilized area stays nearly constant, but the performance of the design improves. On the other hand, by integrating the narrow segments into a wide segment, bitwidth reduces the flexibility and possibility of establishing different connections simultaneously. 
As a case study to inspect probable communication defects, we have implemented a video application with a VGA controller running at 25Mhz for normal 640x480 VGA. A color generator module (CG) communicates with the VGA controller (VC). The color generator gets the X and Y coordinates of the current pixel position from the VGA module, computes the color to be placed at that position and sends it back to the VGA module, which displays the color at the corresponding position. The color generator application is a nice method to detect changes in the communication, becasue this will directly have a visual effect on the screen. The X- and Y-positions are each 12 bits wide and the color is 24 bits wide. This application works well and without communication problems.

\begin{table*}[!htb]
\begin{center}
\begin{tabular}{|l||l||l||l||l||l|}
\hline
DataWidth(bit) $w$& \multicolumn{1}{c|}{\scriptsize{Slices used\#}} & \multicolumn{1}{c|}{\scriptsize{Slices Used\%}} & \multicolumn{1}{c|}{\scriptsize{4-input LUTs used\#}} & \multicolumn{1}{c|}{\scriptsize{4-input LUTs used\%}} & \multicolumn{1}{c|}{\scriptsize{Max frequency(MHz)}}\\
 \hline
\hline
1& \multicolumn{1}{r|}{1367} & \multicolumn{1}{r|}{4} & \multicolumn{1}{r|}{2074} & \multicolumn{1}{r|}{3} & \multicolumn{1}{r|}{105} \\
\hline
8& \multicolumn{1}{r|}{2100} & \multicolumn{1}{r|}{6} & \multicolumn{1}{r|}{3856} & \multicolumn{1}{r|}{4} & \multicolumn{1}{r|}{103} \\
\hline
16& \multicolumn{1}{r|}{3407} & \multicolumn{1}{r|}{10} & \multicolumn{1}{r|}{6108} & \multicolumn{1}{r|}{9} & \multicolumn{1}{r|}{99} \\
\hline
32& \multicolumn{1}{r|}{5084} & \multicolumn{1}{r|}{15} & \multicolumn{1}{r|}{9502} & \multicolumn{1}{r|}{14} & \multicolumn{1}{r|}{94} \\
\hline
\end{tabular}
  \caption[]{Area overhead and performance of 1-dimensional RMBoC with $n=4$ modules and $k=4$ segments per module.}
  \label{res-1d}
\end{center}
\end{table*}

We have also investigated the characteristics of 2-dimensional RMBoC, and the results are presented in Table \ref{res-2d}. The area overhead seems to be too large (more than 50\% of the FPGA area) for practical use, but the maximum frequency is still high (85-96 MHz). Actually for using 2-dimensional circuit routing in future FPGA, the online routing should be done in an additional layer, therefore the area overhead will not be a bottleneck.

\begin{table*}[!htb]
\begin{center}
\begin{tabular}{|l||l||l||l||l||l|}
\hline
DataWidth(bit) $w$& \multicolumn{1}{c|}{\scriptsize{Slices used\#}} & \multicolumn{1}{c|}{\scriptsize{Slices Used\%}} & \multicolumn{1}{c|}{\scriptsize{4-input LUTs used\#}} & \multicolumn{1}{c|}{\scriptsize{4-input LUTs used\%}} & \multicolumn{1}{c|}{\scriptsize{Max frequency(MHz)}}\\
 \hline
\hline
8& \multicolumn{1}{r|}{17192} & \multicolumn{1}{r|}{50} & \multicolumn{1}{r|}{32433} & \multicolumn{1}{r|}{48} & \multicolumn{1}{r|}{96} \\
\hline
16& \multicolumn{1}{r|}{26762} & \multicolumn{1}{r|}{61} & \multicolumn{1}{r|}{37607} & \multicolumn{1}{r|}{56} & \multicolumn{1}{r|}{91} \\
\hline
32& \multicolumn{1}{r|}{28156} & \multicolumn{1}{r|}{83} & \multicolumn{1}{r|}{53872} & \multicolumn{1}{r|}{79} & \multicolumn{1}{r|}{85} \\
\hline
\end{tabular}
  \caption[]{Area overhead and performance of 2-dimensional RMBoC with $n=16$ modules and $k=4$ segments per module and direction.}
  \label{res-2d}
\end{center}
\end{table*}

\section{Conclusion} \label{concl}
In this paper, we have investigated online circuit routing, in particular for dynamic reconfigurable devices. As a practical solution for Xilinx FPGAs, we propose a RMBoC network that has a low area overhead and works with high frequencies. This solution has been implemented; it works completely on Xilinx FPGAs. In addition, we have extended the RMBoC concept to a 2-dimensional one, at the expense of a considerable amount of area. On the other hand, this 2-dimensional network yields a high performance, which can be useful for future generations of FPGAs.

\bibliographystyle{ieee}
\bibliography{rsp05}

\begin{thebibliography}{10}

\bibitem{ABFTV04}
A.~Ahmadinia, C.~Bobda, S.~Fekete, J.~Teich, and J.~van~der Veen.
\newblock Optimal routing-conscious dynamic placement for reconfigurable
  devices.
\newblock In {\em Field-Programmable Logic and Applications, International
  Conference FPL}, pages 847--851, 2004.

\bibitem{baz}
K.~Bazargan, R.~Kastner, and M.~Sarrafzadeh.
\newblock {Fast Template Placement for Reconfigurable Computing Systems}.
\newblock {\em In IEEE Design and Test - Special Issue on Reconfigurable
  Computing}, January-March:68--83, 2000.

\bibitem{micheli}
L.~Benini and G.~D. Micheli.
\newblock {Networks on Chip: A New SoC Paradigm}.
\newblock In {\em IEEE Computer, Vol. 35, NO. 1}, pages 70 -- 80, Jan. 2002.

\bibitem{blodget}
B.~Blodget, C.~Bobda, M.~H{ü}bner, and A.~Niyonkuru.
\newblock {Partial and Dynamically Reconfiguration of Xilinx Virtex-II FPGAs}.
\newblock In {\em Field-Programmable Logic and Applications, International
  Conference FPL}, pages 801--810, 2004.

\bibitem{dynoc}
C.~Bobda, M.~Majer, D.~Koch, A.~Ahmadinia, and J.~Teich.
\newblock {A Dynamic NoC Approach for Communication in Reconfigurable Devices}.
\newblock In {\em Field-Programmable Logic and Applications, International
  Conference FPL}, pages 1032--1036, 2004.

\bibitem{dehon}
A.~DeHon, R.~Huang, and J.~Wawrzynek.
\newblock {Hardware-Assisted Fast Routing}.
\newblock In {\em Proceedings of the IEEE Symposium on Field-Programmable
  Custom Computing Machines}, pages 205--215, Apr. 2002.

\bibitem{elgindy}
H.~A. ElGindy, A.~K. Somani, H.~Schroeder, H.~Schmeck, and A.~Spray.
\newblock {RMB - A Reconfigurable Multiple Bus Network}.
\newblock In {\em Proceedings of the Second International Symposium on
  High-Performance Computer Architecture (HPCA-2)}, pages 108--117, Feb. 1996.

\bibitem{jantsch}
A.~Hemani, A.~Jantsch, S.~Kumar, A.~Postula, J.~Oberg, M.~Millberg, and
  D.~Lindqvist.
\newblock {Network on chip: An architecture for billion transistor era}.
\newblock In {\em Proceeding of the IEEE NorChip Conference}, pages 166--173,
  Nov. 2000.

\bibitem{vernalde}
T.~Marescaux, A.~Bartic, D.~Verkest, S.~Vernalde, and R.~Lauwereins.
\newblock {Interconnection networks enable fine-grain dynamic multitasking on
  FPGAs}.
\newblock In {\em Field-Programmable Logic and Applications, International
  Conference FPL}, pages 795--805, 2002.

\bibitem{pathfinder}
L.~McMurchie and C.~Ebeling.
\newblock {PathFinder: a negotiation-based performance-driven router for
  FPGAs}.
\newblock In {\em Proceedings of the 1995 ACM third international symposium on
  Field-programmable gate arrays}, pages 111--117, Feb. 1995.

\end{thebibliography}
\end{document}